\documentclass[showpacs]{revtex4}
\usepackage{amssymb}
\usepackage{amsmath}
\usepackage{bm}

\usepackage{graphics}

\newcommand{\out}{\operatorname{out}}
\newcommand{\n}{\operatorname{in}}
\begin{document}

\title{Inertia of Intrinsic Spin}

\author{Bahram Mashhoon}

\email{mashhoonb@missouri.edu}
\affiliation{Department of Physics and Astronomy, University of Missouri-Columbia, Columbia,
Missouri 65211, USA}
\author{Helmut Kaiser}
\email{helkaise@indiana.edu}
\affiliation{Indiana University Cyclotron Facility, University of Indiana, Bloomington,
Indiana 47408,
USA}

\begin{abstract}
The state of a particle in space and time is characterized by its mass and spin, which
therefore determine the inertial properties of the particle. The coupling of intrinsic
spin with rotation is examined and the corresponding inertial effects of intrinsic spin
are studied. An experiment to measure directly the spin-rotation coupling via neutron
interferometry is analyzed in detail.
\end{abstract}
\keywords{Spin-rotation coupling\sep Neutron
interferometry}
\pacs{03.30.+p;
03.75.Dg}

\maketitle

\section{Introduction}
\label{s1}

In classical physics any inertial force acting on a particle is necessarily proportional
to the particle's inertial mass. The principle of equivalence of inertial and
gravitational masses would then suggest that inertial forces may be of gravitational
origin---an idea (``Mach's principle") that played a crucial role in the development of
general relativity~\cite{1}. On the basis of the modern geometric interpretation of
Einstein's theory of gravitation, however, the notion that inertial effects are of
gravitational origin must be rejected~\cite{2,3}. Moreover, in quantum theory the
inertial properties of a particle are determined by its inertial mass as well as spin.
Mass and spin characterize the irreducible unitary representations of the inhomogeneous
Lorentz group~\cite{4}. The inertial properties of mass are well known~\cite{5};
therefore, this paper is devoted to the inertial properties of intrinsic
spin.

Spin-rotation coupling and its consequences are described in sections~\ref{s2}
and \ref{s3}. An experiment to measure this effect for neutrons is discussed in
section~\ref{s4}.  

\section{Spin-rotation coupling}
\label{s2}

It follows from the treatment of Dirac's equation in accelerated frames of reference
that in the nonrelativistic regime the Hamiltonian of a Dirac particle contains
$-\bm{\Omega}\cdot \bm{J}$, where $\bm{\Omega}$ is the rotation frequency of the frame
and $\bm{J}=\bm{L}+\bm{S}$ is the total angular momentum of the particle~\cite{6}. The
appearance of $\bm{J}$ is natural, since in quanum theory the total angular momentum is
the generator of rotations. The term $-\bm{\Omega}\cdot \bm{L}$ involving the orbital
angular momentum $\bm{L}$ is responsible for the Sagnac effect, first demonstrated
experimentally for slow neutrons by Werner et al.~\cite{7}. We are interested in the
spin-rotation coupling term $-\bm{\Omega}\cdot \bm{S}$, which does {\em not} depend upon
the inertial mass of the particle; indeed, it is an inertial effect of intrinsic
spin~\cite{8}. It originates from the tendency of intrinsic spin to keep its aspect with
respect to a global background inertial frame (``inertia of intrinsic spin"). From the
standpoint of observers at rest in the rotating frame, the intrinsic spin therefore
precesses in a sense that is opposite to the sense of rotation of the frame. This
precessional motion can be described by the Heisenberg equation of motion with the
Hamiltonian $-\bm{\Omega}\cdot \bm{S}$. Moreover, it can be shown that the coupling of
spin with rotation is fully relativistic~\cite{9}.

To illustrate the general nature of this coupling, consider an observer rotating
uniformly with frequency $\Omega$ about the direction of propagation of a plane
electromagnetic wave of frequency $\omega$. It follows from the Fourier analysis of the
electromagnetic field measured by the rotating observer that the frequency of the wave
in the rotating frame is given by
\begin{equation}\label{eq1} \omega '=\gamma (\omega \mp\Omega),\end{equation}
where $\gamma$ is the Lorentz factor of the observer. Here the upper (lower) sign refers
to positive (negative) helicity radiation. Multiplication by $\hbar$ results in
$E'=\gamma (E-\bm{\Omega}\cdot \bm{S})$, where $\bm{S}$ is the spin of the photon. This
helicity contribution to the Doppler effect has been verified via the GPS~\cite{10}; in
fact, the helicity-rotation coupling is responsible for the phenomenon of {\em phase
wrap-up} \cite{10}. An intuitive explanation of equation~\eqref{eq1} involves the fact
that in a positive (negative) helicity wave, the electromagnetic field rotates with
frequency $\omega\; (-\omega)$ about the direction of propagation of the wave. Thus the
rotating observer perceives positive (negative) helicity radiation with the
electromagnetic field rotating with frequency $\omega -\Omega$ ($-\omega-\Omega$) about
the direction of wave propagation. In addition, the Lorentz factor in equation~\eqref{eq1}
is due to time dilation. More generally, the energy of a particle measured by an
observer rotating with frequency $\Omega$ is given by
\begin{equation}\label{eq2} E'=\gamma (E-\hbar M\Omega),\end{equation}
where $M$ is the total (orbital plus spin) ``magnetic" quantum number along the axis of
rotation ($M=0,\pm 1,\pm 2,\dots $, for a scalar or a vector particle, while $M\mp
\frac{1}{2} =0,\pm 1,
\pm
2,\dots$, for a Dirac particle). For fermions, there exists at
present only indirect evidence for the existence of spin-rotation coupling~\cite{11,12}.

\section{Energy
shift}\label{s3}

\begin{figure}
{\includegraphics{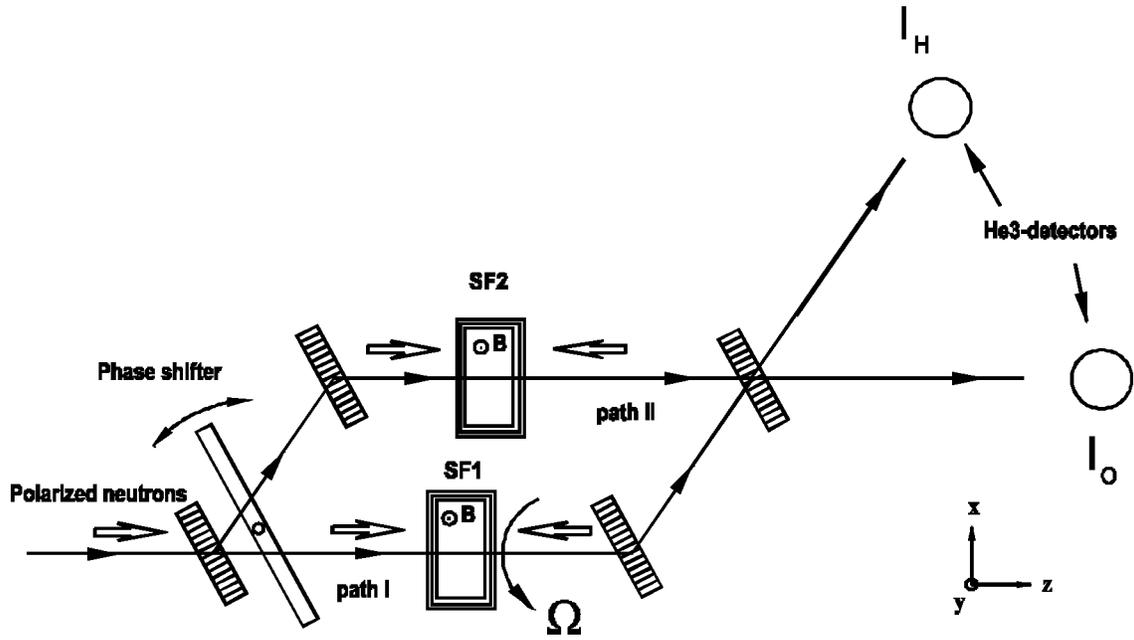}}
\caption{Layout of the proposed experiment.  Longitudinally polarized
neutrons pass through a LLL-interferometer with a static $\pi $-spin flipper SF2 in path
II,
while there is a slowly rotating $\pi$-spin flipper SF1 in path I, having angular velocity
$\bm{\Omega}$ (case
(1)) [or replaced by a static spin flipper with a rotating magnetic field $\bm{B}$
(quadrature coil) in path I (case (2))].\label{f1}}
\end{figure}

 An interesting consequence of the spin-rotation coupling is the energy
shift that would be induced when polarized radiation passes through a
rotating spin flipper. To illustrate this effect, imagine positive helicity
electromagnetic radiation of frequency $ \omega_{\n} $ that is normally incident on
a uniformly rotating half-wave plate. It follows from equation~\eqref{eq1} that
within the half-wave plate $\omega' \approx  \omega_{\n}  - \Omega$, where $\Omega$ is the
rotation frequency and we have assumed that $\gamma \approx 1$. The spacetime in a
uniformly rotating system is stationary; therefore, $\omega'$ remains constant
throughout the rotating half-wave plate. The radiation that emerges has
negative helicity and frequency $ \omega_{\out}$, where $\omega' \approx \omega_{\out} + \Omega$
by equation~\eqref{eq1}. Thus $\omega_{\out} - \omega_{\n}  \approx - 2 \Omega$, so that the photon
energy is down-shifted by $- 2 \hbar \Omega$. Such an energy shift was first
discovered experimentally in the microwave regime~\cite{13}. It follows from the
general nature of spin-rotation coupling that an up/down energy shift
\begin{equation} \delta H \approx - 2\bm{\Omega }\cdot \bm{S} \label{eq3}\end{equation}
occurs whenever spinning particles pass through a $\pi$-spin flipper. An
experiment to measure this effect for slow neutrons was first suggested in~\cite{14}. This concept is elaborated in the rest of this paper. We expect that a
longitudinally polarized neutron passing through a rotating $\pi$-spin flipper
coil would experience an energy shift of $\hbar \Omega$ that would result in a
beat phenomenon in a neutron interferometer as in Figure~\ref{f1}.

    A static $\pi$-spin flipper provides a region of constant uniform magnetic
field $\bm{B}$ over a definite width $w$ such that the interaction $H = - \bm{\mu}_n \cdot \bm{B}$
induces a reversal of the spin direction. Here $\bm{\mu}_n = \gamma_n \bm{S}$ is the
neutron magnetic dipole moment, $\gamma_n$ is the gyromagnetic ratio and $w \approx \pi
v_n / ( |\gamma_n |B )$ , where $v_n$ is the neutron speed. For instance, in Figure~\ref{f1},
$\bm{B} = B \hat{\bm{y}}$ and the
outgoing neutron
has a spin that is antiparallel to the direction of the incident longitudinally polarized neutron. Physically, we
expect that a uniformly rotating spin flipper with $\Omega \ll |\gamma_n |B$ would
be essentially equivalent to a static coil with uniform but variable
magnetic field $\bm{B} = B ( - \hat{\bm{x}} \sin \Omega t + \hat{\bm{y}} \cos \Omega t )$.

\section{Discussion}\label{s4}

We propose a neutron interferometry
experiment using polarized neutrons to be carried out for two equivalent
cases:\vspace{7pt}

\noindent (1) Rotating $180^\circ$- spin flipper coil:

The experimental arrangement is shown in Figure~\ref{f1}.  Neutrons, longitudinally polarized,
pass through our interferometer with $180^\circ$ - spin flipper coils (producing a uniform
static
magnetic field normal to the polarization axis of the neutrons), placed in each of the
two coherent neutron beam paths.  SF1 and SF2 are both optimized such that they flip the
neutron spin by $180^\circ$.  The phase shifter allows us
to set the intensity at
maximum when both coils are aligned parallel.  Keeping the interferometer stationary in
the inertial frame of the laboratory, we then rotate SF1 with angular velocity
$\bm{\Omega}$
parallel to the neutron wave vector and the resulting phase shift can be measured.  After
superposition of the two coherent beams, we expect to observe a sinusoidal intensity
modulation (beating that arises from the spin-rotation coupling) with a beating period
corresponding to $\Omega$, the rate of the rotation of SF1.  In the actual experiment, we
will rotate SF1 approximately $120^\circ$ forth and back, since continuous full rotation
would be
impractical because of wires leading to the power supply and the cooling lines attached to
the spin flipper coil. \vspace{7pt}

\noindent (2) Quadrature coil:

The experimental arrangement is the same as in Figure~\ref{f1},  except that the rotating
$180^\circ$-spin
flipper coil SF1 is replaced by a stationary quadrature coil, which produces a rotating
magnetic field $\bm{B} = (-B\sin\Omega t, B \cos\Omega t, 0)$ normal to the polarization
axis of the neutrons. The expected result should be the same, since the neutron cannot distinguish if SF1 is
physically rotated or the magnetic field is rotated.  The proposed experiment, as
described in case (1), is related to the neutron beat frequency measured by Badurek et
al.~\cite{15} via the Larmor theorem; thus $\bm{\Omega}\Leftrightarrow \gamma_n \bm{B}'$, 
where $\gamma_n$ is the neutron gyromagnetic ratio and $\bm{B}'$ is the magnetic field of
their experiment.

Contemplating an experimental arrangement similar to that employed by
Badurek et al.~\cite{15}, the rotation of the $\pi$-spin flipper with a period 
of about one minute would result in an energy difference of $\sim 10^{-16}$ eV between the two beams of our interferometer. Using neutrons of wavelength 
$\sim$ 2 \AA, we would expect an intensity modulation amplitude of $\sim 30$ counts per
second~\cite{15}. It is interesting to note that it may be possible 
to use magnetized foils~\cite{16} as $\pi$-flippers instead of some of the 
coils discussed in this section.

The proposed experiment  will provide a direct interferometric test of spin-rotation coupling for fermions.  The phenomenon of spin-rotation coupling is of basic
interest since it reveals the rotational inertia of intrinsic spin.

\section*{Acknowledgments}
    
   We are grateful to Sam Werner for many valuable discussions. The work
of H. Kaiser has been supported by the U.S. National Science Foundation
under grant no. PHY-0245679.

\end{document}